\documentclass[aps,twocolumn,groupedaddress]{revtex4}
\usepackage{graphics}
\begin{document}

\markboth{T. Qureshi, P. Chingangbam, S. Shafaq}
{Understanding Ghost Interference}

\title{Understanding Ghost Interference}
\author{Tabish Qureshi}
\address{Center for Theoretical Physics,
Jamia Millia Islamia, New Delhi 110025, India.}
\email{tabish@ctp-jamia.res.in}
\author{Pravabati Chingangbam}
\address{Indian Institute of Astrophysics,
Koramangala, Bangalore-560034, India.}
\email{prava@iiap.res.in}
\author{Sheeba Shafaq}
\address{Department of Physics,
Jamia Millia Islamia, New Delhi 110025, India.}
\email{shafaqsheeba1@gmail.com}

\begin{abstract}
The ghost interference observed for entangled photons is theoretically
analyzed using wave-packet
dynamics. It is shown that ghost interference is a combined effect of
virtual double-slit creation due to entanglement, and quantum erasure of
which-path information for the interfering photon. For the case where the 
two photons are of different color, it is shown that fringe width of the
interfering photon depends not only on its own wavelength, but also on the
wavelength of the other photon which it is entangled with.
\end{abstract}

\keywords{Entanglement; Nonlocality; Ghost interference.}

\maketitle

\section{Introduction}

Quantum entanglement is a concept which has intrigued people since 
the time Einstein, Podolsky and Rosen\cite{epr} first raised some objections
against quantum mechanics, which led Schr\"odinger\cite{entanglement} to
formalize the concept.  Entangled quantum systems, even
though far separated in space, show certain correlation in their
measurement results. Einstein believed that such correlations implied a
nonlocal {\em action at a distance}. Measurement on one system seems to
affect a far away system which it is entangled to. Schr\"odinger went 
further to say, ``{It is rather discomforting that the theory should
allow a system to be steered or piloted into one or the other type of
state at the experimenter's mercy in spite of his having no access to
it."}\cite{entanglement}

As experimental techniques progressed, the effect of entanglement was
experimentally demonstrated. One of the most dramatic demonstrations of
nonlocality is in the so-called ghost interference experiment by
Strekalov et.al.\cite{ghostexpt} In the following we briefly describe
the ghost interference experient. 
\begin{figure}[pb]
\centerline{\resizebox{10.0cm}{!}{\includegraphics{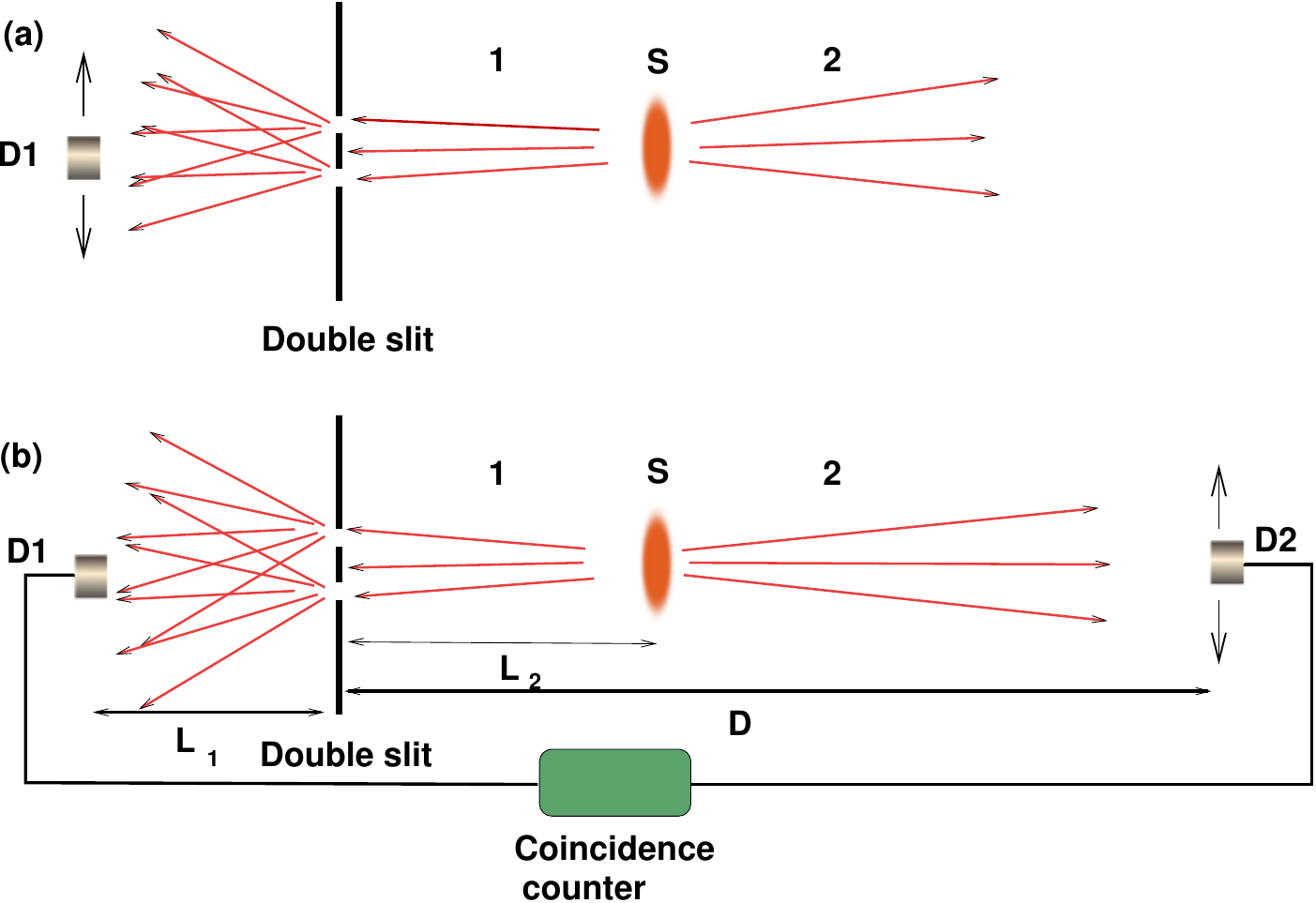}} }
\vspace*{8pt}
\caption{Schematic diagram of the two-slit ghost interference experiment.
Entangled photons 1 and 2 emerge from the SPDC source S and travel in
different directions along x-axis.}
\label{ghostexpt}
\end{figure}
Two entangled photons emerge from a spontaneous parametric down-conversion
(SPDC) source S. Photon 1 and 2 are made to move in different directions,
by a beam-splitter. A double-slit is kept in the path of photon 1, and
there is a scanning detector D1 behind it (see Fig. \ref{ghostexpt}(a)). No first order interference is
observed for photon 1. This is surprising because normally
one expects Young's double-slit interference. The second surprise of the
experiment is that when photon 2 is observed by detector D2,
{\em in coincidence} with a {\em fixed} detector D1 detecting photon 1,
photon 2 shows a double-slit interference pattern (see Fig. \ref{ghostexpt}(b)).
Note that photon 2 does not pass through any double-slit.
The third surprise of the experiment is that the fringe width of the 
interference pattern follows a Young's interference formula
$w = {\lambda D\over d}$, where $D$ is a very curious distance from double-slit,
right throught the SPDC source $S$ to the detector D2. Note that photon 2
does not even travel that much of distance.

To explain the experimental results, the authors presented a geometrical
model where the entangled photons, starting from a spatially extended source,
travelled in exactly oppositely directed paths. The path length depends
on the point of origin of the photon pair in the extended source. The
difference in the lengths of two such paths passing through different slits
was shown to lead to ghost interference.\cite{ghostexpt} The absence of
first order interference behind the double-slit was attributed to ``the
considerably large angular propagation uncertainty of a single SPDC
photon.\cite{ghostexpt} The experiment attaracted a lot of debate and
research attention.\cite{ghostimaging,rubin,zhai,jie,zeil2,pravatq}

An explanation based on geometric trajectories is not very satisfying, and
one would like to look for a more quantum argument. Also, from the explanation
provided, the origin of ghost interference and the absence of first order
interference seem to be not connected. In the following we will carry out
an analysis based on time evolution of quantum wave-packets, and show that
it gives us a much better understanding of ghost interference. In our analysis,
we will show that the origin of ghost interference and the absence of first
order interference, have a much deeper connection.

Recently a modified ghost interference experiment has been carried out
using photon pairs generated via spontaneous four-wave mixing
(SFWM).\cite{twocolor} This experiment is novel in the aspect that 
the correlated photons in a pair are of different
color, with wavelengths $\lambda_1=1530$ nm and $\lambda_2=780$ nm. This
phenomenon was called {\em two-color ghost interference} by the authors.
We will analyze this experiment too, and present some interesting predictions
for it.
 
\section{Wave-Packet Analysis}

\subsection{The entangled state}

In order to theoretically analyze ghost interference, a
major challenge is to formulate a well-behaved entangled state which
captures the essence of entangled photons. The so-called EPR state is
not well behaved in the sense that it is like a Dirac delta function. We
propose, what we call, a generalized EPR state as follows:
\begin{equation}
\Psi(z_1,z_2) = C\!\int_{-\infty}^\infty dp
e^{-p^2/4\hbar^2\sigma^2}e^{-ipz_2/\hbar} e^{i pz_1/\hbar}
e^{-{(z_1+z_2)^2\over 4\Omega^2}}, \label{state}
\end{equation}
where $C$ is a normalization constant, and $\sigma,\Omega$ are certain
parameters whose physical significance will become clear in the following. The
state (\ref{state}), unlike the EPR state, is well behaved and fully
normalized. In the limit $\sigma,\Omega\to \infty$ the state (\ref{state})
reduces to the EPR state.

The photons of the pair are assumed to be travelling in opposite directions
along the x-axis, but the entanglement is in the z-direction. In our analysis
we will ignore the dynamics along the x-axis as it does not affect entanglement.
We just assume that during evolution for a time $t$, the photon travels a
distance equal to $ct$.  Integration over $p$ can be performed in 
(\ref{state}) to obtain:
\begin{equation}
\Psi(z_1,z_2) = \sqrt{ {\sigma\over \pi\Omega}}
 e^{-(z_1-z_2)^2\sigma^2} e^{-(z_1+z_2)^2/4\Omega^2} .
\label{psi0}
\end{equation}
The uncertainty in position and the wave-vector of the two photons,
along the z-axis, is given by
\begin{eqnarray}
\Delta z_1 &=& \Delta z_2 = \sqrt{\Omega^2+1/4\sigma^2},\nonumber\\
\Delta k_{1z} &=& \Delta k_{2z} = 
{1\over 2}\sqrt{\sigma^2 + {1\over 4\Omega^2}}~. \label{unc}
\end{eqnarray}
For the above it is clear that $\Omega$ and $\sigma$ quantify the position
and momentum spread of the photons in the z-direction.

\subsection{Time evolution}

We will first lay out our strategy for time evolution of a photon
wave-packet. If the state at time $t=0$ is $\psi(z,0)$, the state at
a later time is given by
\begin{equation}
\psi(z,t) = {1\over 2\pi}\int_{\-\infty}^{\infty}
\exp(ik_zz - i\omega(k_z)t) \tilde{\psi}(k_z,0) dk_z ,
\label{ppsit1}
\end{equation}
where $\tilde{\psi}(k_z,0)$ is the Fourier transform of $\psi(z,0)$ with
respect to $z$. Now photon is approximately travelling in the x-direction,
but can slightly deviate in the z-direction (which allows it to pass through
slits located at different z-positions), so that its true wave-vector will
have a small component in the z-direction too. So
\begin{equation}
\omega(k_z) = c\sqrt{k_x^2 + k_z^2}
\end{equation}
Since the photon is travelling along x-axis by and large, we can write
$k_x \approx k_0$, where $k_0$ is the wavenumber of the photon associated
with its wavelength, $k_0 = 2\pi/\lambda$. The dispersion along z-axis can
then be approximated by
\begin{equation}
\omega(k_z) \approx ck_0 + ck_z^2/2k_0 .
\end{equation}
Using this, Eqn. (\ref{ppsit1}) becomes
\begin{equation}
\psi(z,t) = {e^{ick_0t}\over 2\pi}\int_{\-\infty}^{\infty}
\exp(ik_zz - ictk_z^2/2k_0) \tilde{\psi}(k_z,0) dk_z
\label{ppsit2}
\end{equation}

Coming back to our problem of entangled photons,
we assume that after travelling for a time $t_0$, photon 1 reaches the 
double slit ($ct_0 = L_2$), and photon 2 travels a distance $L_2$ towards
detector D2.
Using the strategy outlined in the preceding discussion, we can write
the state of the entangled photons after a time $t_0$ as follows:
\begin{eqnarray}
\psi(z_1,z_2,t_0) &=& {e^{2ick_0t_0}\over 4\pi^2}\int_{\-\infty}^{\infty} dk_1
\exp(ik_1z_1 - ict_0k_1^2/2k_0)\nonumber\\
&&\int_{\-\infty}^{\infty} dk_2
\exp(ik_2z_2 - ict_0k_2^2/2k_0) \tilde{\psi}(k_1,k_2,0) ,\nonumber\\
\label{psit0}
\end{eqnarray}
where $\tilde{\psi}(k_1,k_2,0)$ is the Fourier transform of (\ref{psi0}) with
respect to $z_1,z_2$.

\subsection{Effect of double-slit}

In order to see the effect of the double-slit on the entangled state, one
would normally model a potential for the double-slit, and calculate the
evolution of the state in that potential. That is not an easy task. We
will follow an alternative strategy which captures the essence of the
effect of the double-slit on the state, without going into tedious
calculation. When the state interacts with a single-slit, let us assume that
what emerges from a single slit is a Gaussian wave-packet centered at the
location of the slit, and whose width is related to the width of the slit.
So, if the two slits are A and B, the packets which
pass through will be, say, $|\phi_A(z_1)\rangle$ and $|\phi_B(z_1)\rangle$,
respectively. Some part of the state of particle 1 will get blocked. Let us
represent it by $\chi(z_1)$. These three states are obviously orthogonal,
and the actual state of particle 1 can be expanded in this basis. 
\begin{equation}
|\Psi(z_1,z_2,t_0)\rangle = |\phi_A\rangle\langle\phi_A|\Psi\rangle
+ |\phi_B\rangle\langle\phi_B|\Psi\rangle +
|\chi\rangle\langle\chi|\Psi\rangle . \label{slit}
\end{equation}
The terms $\langle\phi_A|\Psi\rangle, \langle\phi_B|\Psi\rangle,
\langle\chi|\Psi\rangle$ are states of particle 2 and can be explicitly
calculated as
\begin{eqnarray}
\psi_A(z_2) &=& \langle\phi_A(z_1)|\Psi(z_1,z_2,t_0)\rangle \nonumber\\
\psi_B(z_2) &=& \langle\phi_B(z_1)|\Psi(z_1,z_2,t_0)\rangle \nonumber\\
\psi_\chi(z_2) &=& \langle\chi(z_1)|\Psi(z_1,z_2,t_0)\rangle \label{psi}
\end{eqnarray}
So, the entangled state we get after particle 1 crosses the double-slit is:
\begin{equation}
|\Psi(z_1,z_2)\rangle = |\phi_A\rangle|\psi_A\rangle
+ |\phi_B\rangle|\psi_B\rangle +
|\chi\rangle|\Psi_\chi\rangle ,
\end{equation}
where $|\phi_A\rangle$ and $|\phi_B\rangle$ are states of particle 1,
and $|\psi_A\rangle$ and $|\psi_B\rangle$ are states of particle 2.
The first two terms represent the amplitudes of particle 1 passing through
the slits, and the last term represents the amplitude of it getting
reflected or blocked. The linearity of the Schr\"odinger equantion assures that
the first two terms and the last term evolve independently. Since the experiment
only looks for those photons 1 which have passed through the double-slit,
we might as well throw away the last term. Doing this will not change anything
except for renormalizing the part of the state that we retain.

In the following, we assume that $|\phi_A\rangle$, $|\phi_B\rangle$, are
Gaussian wave-packets:
\begin{eqnarray}
\phi_A(z_1) &=& {1\over(\pi/2)^{1/4}\sqrt{\epsilon}} e^{-(z_1-z_0)^2/\epsilon^2}
,\nonumber\\
\phi_B(z_1) &=& {1\over(\pi/2)^{1/4}\sqrt{\epsilon}} e^{-(z_1+z_0)^2/\epsilon^2},
\end{eqnarray}
where $\pm z_0$ is the z-position of slit A and B, respectively, and $\epsilon$
their widths. Thus, the distance between the two slits is $2 z_0 \equiv d$.

Using (\ref{psi}) and (\ref{psit0}), wavefunctions for $|\psi_A\rangle$,
$|\psi_B\rangle$ can be calculated, which, after normalization, have the form
\begin{equation}
\psi_A(z_2) = C_2 e^{-{(z_2 - z_0')^2 \over \Gamma}},~~~
\psi_B(z_2) = C_2 e^{-{(z_2 + z_0')^2 \over \Gamma}} ,
\end{equation}
where $C_2 = (2/\pi)^{1/4}(\sqrt{\Gamma_R} + {i\Gamma_I\over\sqrt{\Gamma_R}})^{-1/2}$,
\begin{equation}
z_0' = {z_0 \over {4\Omega^2\sigma^2+1\over
4\Omega^2\sigma^2-1} + {4\epsilon^2 \over 4\Omega^2-1/\sigma^2}},
\end{equation}
and
$
\Gamma = \frac{\epsilon^2 + {1\over\sigma^2}(1+{\epsilon^2\over 4\Omega^2})
 + {ict_0\over 2\pi\sigma^2\Omega^2}{\lambda_1\lambda_2\over\lambda_1+\lambda_2} +{ict_0\over\pi}(\lambda_1+\lambda_2)(1+{1\over 4\Omega^2\sigma^2})}
{1 + {\epsilon^2\over\Omega^2}+{ict_0\over 4\pi\Omega^2}(\lambda_1+\lambda_2
+{\lambda_1\lambda_2\over\lambda_1+\lambda_2})+ {1\over 4\Omega^2\sigma^2}} .
$
Here $\Gamma_R, \Gamma_I$ are the real and imaginary parts of $\Gamma$,
respectively.

Thus, the state which emerges from the double slit, has the following form
\begin{eqnarray}
\Psi(z_1,z_2) &=& c~e^{-(z_1-z_0)^2/\epsilon^2}
e^{-{(z_2 - z_0')^2 \over \Gamma}}\nonumber\\
&& + c e^{-(z_1+z_0)^2/\epsilon^2}
e^{-{(z_2 + z_0')^2 \over \Gamma}} \label{virtual},
\end{eqnarray}
where $c = (1/\sqrt{\pi\epsilon})(\sqrt{\Gamma_R} +
{i\Gamma_I\over\sqrt{\Gamma_R}})^{-1/2}$. In obtaining this expression, we
have dropped the phase factor in (\ref{psit0}), as it is not important for
our final analysis.
Equation (\ref{virtual}) represents two wave-packets of photon 1,
of width $\epsilon$, and localized at $\pm z_0$, entangled with two
wave-packets of photon 2, of width
${\sqrt{2}|\Gamma|\over\sqrt{\Gamma+\Gamma^*}}$, localized at
$\pm z_0'$.

Even at this early stage one can see that the amplitudes of photon 1
passing through slits A and B are correlated to spatially separated
wave-packets of 
photon 2. Thus, in principle one can detect photon 2 and know which
slit, A or B, photon 1 passed through. By Bohr's principle of complementarity,
if one knows which slit photon 1 passed through, it cannot show any
interference. This is the fundmental reason
for non-observance of first order interference in photon 1 in the ghost
interference experiment.

\subsection{Ghost interference}

The parts of the state for photon 2 are in the form of two spatially
separated wave-packets. This suggests that photon 2 has passed
through a {\em virtual double-slit} (of slit separation $2z_0'$),
conditioned on photon 1 having passed through the real double-slit.
As photon 2 evolves in time, the two wave-packets will overlap, and
interference is a possibility. This is consistent with the ghost imaging
observed for entangled photons.\cite{imaging} It must be emphasized here
however, that although entanglement leads to ghost imaging,
entanglement is not a requirement for it. Classical correlations
in light are enough for demonstrating ghost imaging. This subject has been 
widely debated, and we can only refer the reader to two review articles
and references therein.\cite{shapiro1,shih3}

Before reaching detector D2, particle 2 evolves for a further time $t$.
The time evolution, transforms the state (\ref{virtual}) to
\begin{eqnarray}
\Psi(z_1,z_2,t) &=& 
C_t \exp\left[{{-(z_1-z_0)^2\over\epsilon^2+ict\lambda_1/\pi}}\right]
 \exp\left[{{-(z_2 - z_0')^2 \over \Gamma+ict\lambda_2/\pi}}\right]
\nonumber\\
&&+ C_t \exp\left[{{-(z_1+z_0)^2\over\epsilon^2+ict\lambda_1/\pi}}\right]
 \exp\left[{{-(z_2 + z_0')^2 \over \Gamma+ict\lambda_2/\pi}}\right],\nonumber\\
\label{psifinal}
\end{eqnarray}
where 
\begin{equation}
C(t) = {1\over \sqrt{\pi}\sqrt{\epsilon+ict\lambda_1/\epsilon\pi}
\sqrt{\sqrt{\Gamma_r}+(\Gamma_i+ict\lambda_2/\pi)/\sqrt{\Gamma_r}}}.
\end{equation}
If the correlation because of entanglement between the photons is good,
one can make the following approximation:
$\Omega \gg \epsilon$, $\Omega \gg 1/\sigma$ and $\Omega \gg 1$. In this
limit,
\begin{equation}
\Gamma \approx \gamma^2 + 2i\hbar t_0/\mu,~~~ z_0' \approx z_0,
\end{equation}
where $\gamma^2 = \epsilon^2 + 1/\sigma^2$.

The wave-function (\ref{psifinal}) represents the combined state of the
two photons when they reach detectors D1 and D2 respectively.
The stage is now set to calculate the probability of coincident counting
of D1 located at $z_1$ and D2 located at $z_2$. 

If D1 and D2 are located at $z_1$ and $z_2$ respectively, the
probability density of their coincident count is given by
\begin{eqnarray}
P(z_1,z_2) &=& |\Psi(z_1,z_2,t)|^2 \nonumber\\
&=& |C_t|^2 \left(  
\exp\left[-{2(z_1-z_0)^2\over\epsilon^2+{\lambda_1^2 L_1^2\over\pi^2\epsilon^2}}
-{2(z_2 - z_0)^2 \over \gamma^2+{(\lambda_2 L+\lambda_1 L_2)^2\over\pi^2\gamma^2}}\right]\right. \nonumber\\
&&+ \exp\left[-{2(z_1+z_0)^2\over\epsilon^2+({\lambda_1 L_1\over\pi\epsilon})^2}
-{2(z_2 + z_0)^2 \over \gamma^2+({\lambda_2 L+\lambda_1 L_2\over\pi\gamma})^2}\right]
\nonumber\\
&&+ \exp\left[-{2(z_1^2+z_0^2)\over\epsilon^2+({\lambda_1 L_1\over\pi\epsilon})^2}
-{2(z_2^2 + z_0^2) \over \gamma^2+({\lambda_2 L+\lambda_1 L_2\over\pi\gamma})^2}\right]\nonumber\\
&&\left.\times 2\cos\left[\theta_1 z_1 + \theta_2 z_2\right]\right),
\label{pattern}
\end{eqnarray}
where 
\begin{equation}
\theta_1 = {2d\lambda_1 L_1/\pi\over \epsilon^4 + \lambda_1^2L_1^2/\pi^2},~~~
\theta_2 = {2\pi d[\lambda_2 L+\lambda_1L_2]\over \gamma^4\pi^2 +
(\lambda_2 L+\lambda_1L_2)^2},
\end{equation}
$L = L_1 + L_2$, and
\begin{equation}
C_t = {1\over \sqrt{\pi}\sqrt{\epsilon+{i\lambda_1L_1\over\pi\epsilon}}
\sqrt{\gamma+{i\lambda_2 L+i\lambda_1L_2\over\pi\gamma}}}.
\end{equation}

\section{Results}

\subsection{Ghost interference}

Let us first analyze the original ghost interference experiment when the
entangled photons have the same wave-length $\lambda$, and detector D1 is fixed at $z_1=0$. In that case, (\ref{pattern}) reduces to
\begin{eqnarray}
P(z_1,z_2) &=& |\Psi(z_1,z_2,t)|^2 \nonumber\\
&=& |C_t|^2 \exp\left[-{2z_0^2\over\epsilon^2+{\lambda^2 L_1^2\over\pi^2\epsilon^2}}\right]
\exp\left[
-{2(z_2^2 + z_0^2) \over \gamma_D^2}\right]\nonumber\\
&&2\cosh(4z_2z_0/\gamma_D^2)
\times\left(1 + {\cos\left[\theta_D z_2\right]\over
\cosh(4z_2z_0/\gamma_D^2)}\right),\nonumber\\
\label{pattern-ghost}
\end{eqnarray}
where
\begin{equation}
\theta_D = {2\pi d\lambda D\over \gamma^4\pi^2 +
\lambda^2 D^2},
\end{equation}
and $\gamma_D^2 = \gamma^2 + (\lambda D/\pi\gamma)^2$.
For $\gamma^2\ll \lambda D/\pi$, 
(\ref{pattern-ghost}) represents an interference pattern for photon 2
with the fringe width given by
\begin{equation}
 w_2 \approx {\lambda D\over d}.
\label{young1}
\end{equation}
This is ghost interference, and one should notice that the distance $D$
appearing in the formula is the distance from the double-slit, right through
the source to detector D2. This is exactly what was seen in the experiment.\cite{ghostexpt}

\subsection{Two-color ghost interference}

Now we analyze the two-color ghost interference experiment of
Ding et.al.\cite{twocolor} When the two photons have difference wave-lengths,
and detector D1 is fixed at $z_1=0$, (\ref{pattern}) reduces to
\begin{eqnarray}
P(z_1,z_2) &=& |C_t|^2 \exp\left[-{2z_0^2\over\epsilon^2+
{\lambda^2 L_1^2\over\pi^2\epsilon^2}}\right] \exp\left[
-{2(z_2^2 + z_0^2) \over \gamma_L^2}\right]\nonumber\\
&&2\cosh(4z_2z_0/\gamma_L^2)
\times\left(1 + {\cos\left[\theta_L z_2\right]\over
\cosh(4z_2z_0/\gamma_L^2)}\right),\nonumber\\
\label{pattern-ghost}
\end{eqnarray}
where
\begin{equation}
\theta_L = {2\pi d[\lambda_2 L+\lambda_1L_2]\over \gamma^4\pi^2 +
(\lambda_2 L+\lambda_1L_2)^2},
\end{equation}
and $\gamma_L^2 = \gamma^2 + ({\lambda_2L+\lambda_1L_2\over\pi\gamma})^2$.
The expression (\ref{pattern-ghost}) is an interference pattern for photons 2 which are detected in coincidence with a fixed D1 (see Fig. 2).
For $\gamma^2\ll \lambda_2 L/\pi$,
(\ref{pattern-ghost}) represents an interference pattern for photon 2
with the fringe width given by
\begin{equation}
w_2 \approx {\lambda_2(L_1+L_2)\over d} + {\lambda_1L_2\over d}.
\label{young2}
\end{equation}
One can see that the fringe width of the ghost interference of photon 2 depends
not only on its wavelength $\lambda_2$, it also depends on the wavelength
$\lambda_1$ of the photon which it is entangled with. We believe this is a
highly non-classical feature.

\begin{figure}[h]
\centerline{\resizebox{10.0cm}{!}{\includegraphics{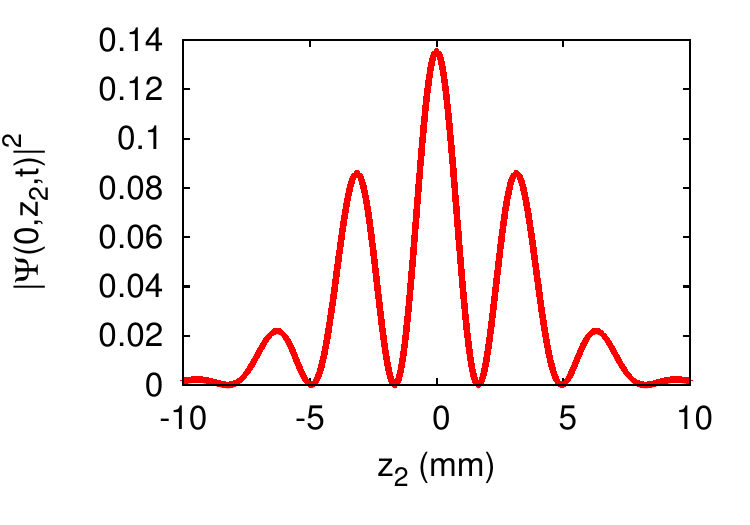}}}
\caption{ Probability density of particle 2 as a function of the position of
detector D2, with D1 fixed at $z_1=0$, for $\lambda_1=1530$ nm,
$\lambda_2=780$ nm, $D=1.8$ m,
$L_1=1.15$ m, $L_2 = 32.5$ cm, $d=0.5$ mm, $\epsilon=0.1$ mm and
$\gamma=0.11$ mm }
\label{graph}
\end{figure}

\subsection{Effect of converging lens}

In the two-color ghost interference experiment, Ding et al.\cite{twocolor}
have used a converging lens between the source and
detector D2. So the fringe-width formula given by (\ref{young2}) doesn't directly
apply. However, if one were to repeat this experiment without the converging
lens, the validity of the formula (\ref{young2}) can be tested.

\begin{figure}[h]
\centerline{\resizebox{10.0cm}{!}{\includegraphics{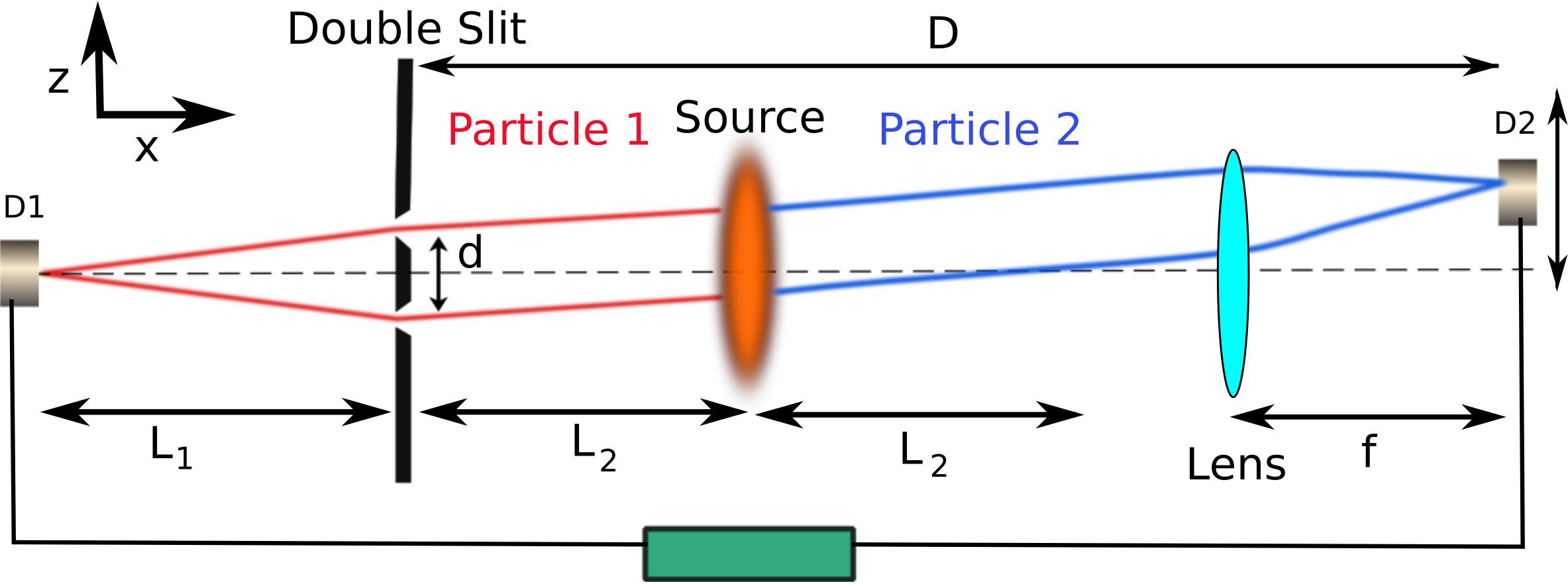}}}
\caption{ The setup for a two-color ghost interference experiment with
a converging lens added. The lens is kept at a distance $f$ before the
detector D2, where $f$ is its focal length. The distance between the
source and the lens is $L_1+L_2-f$.}
\label{lensfig}
\end{figure}

In the following we incorporate the effect of a coverging lens in our
theoretical analysis. This allows
us to make contact with Ding et al.'s experimental results. In order to
do that, we consider an experimental setup similar to that of
Ding et al. \cite{twocolor} (see Fig. \ref{lensfig}). 
A converging lens of focal length $f$ is kept at distance $f$ before 
detector D2. As a result, the distance between the source and the lens
is $L_1+L_2-f$.

The effect of a convegring lens on a general Gaussian wave-packet is 
such that in its subsequent dynamics, it narrows instead of spreading.
In general, we can quantify the effect of the lens by a unitary
transformation of the form\cite{sheebatq}
\begin{eqnarray}
  \mathbf{U}_f {(\pi/2)^{-1/4}\over\sqrt{\sigma+{i\Lambda L\over\sigma}}} 
\exp\left({-z_1^2 \over \sigma^2+i\Lambda L}\right) =
{(\pi/2)^{-1/4}\over\sqrt{{\sigma f\over L-f}-{i\Lambda L\over
\sigma}}} \nonumber\\
\exp\left({-z_1^2 \over \left({\sigma f\over L-f}\right)^2-i\Lambda {fL\over L-f}}\right),
\label{lens}
\end{eqnarray}
where $L$ is the distance the wave-packet, of an initial width $\sigma$, 
traveled before passing through the lens, and $\lambda=\Lambda \pi$ is the
wavelength of the particle. This transformation respects the 
classical thin lens equation in the following way.
If a Gaussian wave-packet of width $\sigma$ starts from a distance
$L$ from the lens, it 
converges back until the imaginary terms in the exponent disappear.
This happens at a distance $u$ such that ${1\over u} = {1\over f} - {1\over L}$.
The width of the wave-packet at that time is ${\sigma f\over L-f}$.

In our calculation without a lens, particles 1 and 2 travel a distance $L_2$
so that photon 1 passes through the double-slit. When the photons emerge
from the double-slit, the two-photon state is given by eqn. (\ref{virtual}).
The situation in this case too remains the same, hence (\ref{virtual})
still holds. However, after this, instead of travelling
a distance $L_1$ to reach D2, photon 2 now travels a distance $L_1-f$
to reach the lens. The two-photon state at this time is given by
\begin{eqnarray}
\Psi(z_1,z_2,t) &=& 
c_1 e^{{{-(z_1-z_0)^2\over\epsilon^2+i\Lambda_1(L_1-f)}}}
 e^{{{-(z_2 - z_0')^2 \over \Gamma+i\Lambda_2(L_1-f)}}}\nonumber\\
&&+ c_1 e^{{{-(z_1+z_0)^2\over\epsilon^2+i\Lambda_1(L_1-f)}}}
 e^{{{-(z_2 + z_0')^2 \over \Gamma+i\Lambda_2(L_1-f)}}},
\label{beforelens}
\end{eqnarray}
where
$c_1 = \left(\sqrt{\pi}\sqrt{\epsilon+i\Lambda_1L_1\epsilon}
\sqrt{\sqrt{\Gamma_r}+(\Gamma_i+i\Lambda_2L_1)/\sqrt{\Gamma_r}}\right)^{-1}$
and $\Gamma \approx \gamma^2 + i(\Lambda_1+\Lambda_2)L_2$.
Photon 2 wave-packet, when it reaches the lens, has {\em apparently} travelled
a distance $(1+\lambda_1/\lambda_2)L_2+L_1$, and has an original width $\gamma$. 

We apply the lens transformation (\ref{lens}) on (\ref{beforelens}) and
let the wave-packets evolve for a further distance $f$ to reach the respective
detectors 1 and 2.
The probability density of finding photon 2 at $z_2$, given that photon 1
is detected at $z_1=0$ , is given by
\begin{eqnarray}
P(0,z_2) &\approx& 2|C_f|^2
e^{-{2z_0^2\over\Delta_1} -{2(z_2^2 + z_0^2) \over \Delta_2}}
\left( 
\cosh\left({4z_2z_0 \over \Delta_2}\right)\right. \nonumber\\
&&\left.+ \cos\left[{2\pi z_2d\over \lambda_2f\left(1+{\alpha L_2+L_1-f\over
\alpha L_2+L_1-2f}\right)}\right]\right),
\label{plot2}
\end{eqnarray}
where
$\Delta_1=\epsilon^2+{\lambda_1^2 L_1^2\over\pi^2\epsilon^2}$,
$\Delta_2=\left({\gamma f\over(1+\alpha)L_2+L_1-2f}\right)^2+
\lambda_2^2\left({2\alpha L_2+2L_1-3f\over
\gamma\pi}\right)^2$,
$\alpha = 1+\lambda_1/\lambda_2$,
and\\
$C_f = {1\over \sqrt{\pi}\sqrt{\epsilon+{i\lambda_1L_1\over\pi\epsilon}}
\sqrt{{\gamma f\over\alpha L_2+L_1-2f}+
i\lambda_2{2\alpha L_2+2L_1-3f\over
\gamma\pi}}}.$
Expression (\ref{plot2}) represents an interference pattern with a fringe-width
given by
\begin{equation}
w_{2} = {\lambda_2 f\left(1+{\alpha L_2+L_1-f\over
\alpha L_2+L_1-2f}\right)\over d}.
\end{equation}
Comparing the above with (\ref{young2}), we see that introducing a converging
lens masks the dramatic effect of
entanglement to large extent. A more simplistic analysis by Ding et al.
gave $w_{2} = {\lambda_2 f\over d}$.\cite{twocolor} 

\section{Discussion}

From the wave-packet analysis carried out here, the following inferences
can be drawn. When one of the two entangled photons (photon 1) passes through a
double-slit, the other one experiences a {\em virtual double-slit} or a
ghost image of the real double-slit, by virtue of their entangled state.
If one is only interested in those photon pairs whose photon 1 has passed 
through the double-slit, the resultant entangled state consists of two
distinct wave-packets of photon 1, correlated with two distinct wave-packets
of photon 2. One can, in principle, detect photon 2 without disturbing 
photon 1, and can figure out which of the two slits photon 1 went through.
Bohr's principle of complementarity implies that if we know which of the two
slits photon 1 went through, it cannot show any interference. This is the
real reason why first order interference behind the double-slit is absent
in Strekalov et al.'s experiment.\cite{ghostexpt}

Since photon 2 experiences a virtual double-slit (conditioned on photon 1
passing through the real double-slit), it has the potential to show an 
interference pattern. However, by virtue of the entangled wave-packet state,
photon 1 also carries information on which of the two virtual slits photon
2 passed through. Thus photon 2 should also not show any interference.
However, in the experiment, detector D1 is fixed at $z_1=0$, far away from
the double-slit. Photon 1 from either of the two slits is equally likely
to reach the fixed detector D1. Thus by the act of fixing D1, we lose
the knowledge of which slit photon 1 went through. The which-way
information about photon 1, and consequently about photon 2, is {\em erased}.
It is well known that if the which-way information in a two-slit experiment
is erased, there are ways in which the interference pattern can be brought
back. This phenomenon goes by the name of {\em quantum erasure}.\cite{marcelo}
So, ghost interference is a result of formation of a virtual double-slit,
by virtue of entanglement, and {\em quantum erasure} of which-way information
for photon 2. A curious feature of this phenomenon is that the virtual
double-slit seems to be located, not anywhere between the source and D2,
as one might naively expect, but exactly at the location of the real
double-slit! Photon 2 goes nowhere near that region. The distance $D$ that
appears in the formula for the fringe-width
of ghost interference (\ref{young1}), is never travelled by photon 2.

As a corollary, if the detector D2 is fixed instead of D1, and D1 is scanned
along z-axis,
photon 1 will show an interference pattern, but only in coincidence with
the fixed detector D2. The fringe-width in this case, will depend on the
actual distance between the double-slit and D1, and not on $D$.

In the case of the two-color ghost interference, the situation is more
interesting. The fringe-width of the ghost interference shown by photon 2
(\ref{young2}),
depends not only on its own wavelength, but also on the wavelength of
photon 1, which it is entangled with. This is a highly non-classical
feature, arising from true entangelment. This effect is hidden in the
case where $\lambda_1 = \lambda_2$. The two-color ghost interference
experiment was carried out in the presence of a converging lens. Our analysis
shows that the converging lens masks the dramatic feature seen in
(\ref{young2}) to a large extent. Formula (\ref{young2}) can be verified
if the two-color ghost interference experiment is carried out without the
converging lens.

\section*{Acknowledgments}

Tabish Qureshi thanks the organisers of the International Program on Quantum
Information 2014, IOP, Bhubaneswar, for making possible a lively debate on
various foundational issues.

\vspace*{-6pt}   

\end{document}